\documentstyle[preprint,aps,eqsecnum]{revtex}
\begin{document}
\vskip2cm
\title{Correction Factors
for Reactions involving
\protect$q \protect\bar q$ Annihilation or Production }
\author{
Lali Chatterjee$^{1,2}$ and Cheuk-Yin Wong$^1$}
\address{
${}^1$Oak Ridge National Laboratory, Oak Ridge, TN 37831
}
\address {\rm and }
\address{ ${}^2$Jadavpur University, Calcutta 700032, India.}
\maketitle
\vspace{2cm}
\begin{abstract}
\noindent
In reactions with $q \bar q$ production or $q\bar q$ annihilation,
initial- and final-state interactions give rise to large
corrections to the lowest-order cross sections.  We evaluate the
correction factor first for low relative kinetic energies by studying the
distortion of the relative wave function.  We then follow the
procedure of Schwinger to interpolate this
result with the well-known perturbative QCD vertex correction factors
at high energies, to obtain an explicit semi-empirical correction factor
applicable to the whole range of energies.   The correction
factor predicts an enhancement for $q\bar q$ in color-singlet states
and a suppression for color-octet states, the effect increasing as
the relative velocity decreases.  Consequences on dilepton production
in the quark-gluon plasma, the Drell-Yan process, and heavy quark
production processes are discussed.

\end{abstract}



\narrowtext
\newpage
\section{ Introduction}

      The possibility of nuclear matter going through a phase
transition into a deconfined quark gluon plasma (QGP) state during
high-energy heavy-ion collisions makes these collisions the focus of
intense experimental and theoretical research \cite{Won94}.  In these
collisions, the constituents of the
possible QGP or the partons of the colliding nucleons can react at varying
energies.  Their reaction products such as dileptons and photons
provide the signals and the backgrounds
for the detection of the quark-gluon plasma.   The magnitudes
of the signals depend on the cross sections for dilepton and photon
production from the constituents of the
plasma.  The rates of approach to
chemical equilibrium and thermal equilibrium of the plasma depend also
 on the cross sections for reactions among the constituents.

When quarks, antiquarks and gluons interact, they interact
as
constituents in the quark-gluon plasma or as partons in the colliding nucleon.
However,
the lowest-order Feynman diagrams and the next-to-leading order
Feynman diagrams are the same for
the basic reaction processes involving $q$, $\bar q$ and gluons,
whether the reaction takes place in the
environment of the quark-gluon plasma or in the environment of partons
in nucleon-nucleon collisions.
In these basic reaction of quarks, antiquarks and gluons,
the next-to leading order
diagrams, including the initial- and final-state interactions and gluon
radiations, give rise to large
corrections to the lowest-order cross sections.
It will
be useful to develop an analytical semi-empirical
correction factor for the basic reaction
processes over the whole range of relative energies so that in the
next level of approximation
the basic lowest-order reaction cross sections of quarks, antiquarks and
gluons can be corrected
on the same footing.  Additional effects and refinements
such as the plasma screening and temperature
can be added on in the future as our theoretical
understanding is  developed further.

For  processes involving $q\bar q$ production and annihilation
at high energies, it has been already
generally recognized in perturbative QCD that the lowest-order Feynman
diagrams give only an approximate description
\cite{Fie89}-\cite{Vog91}.
In the case of the Drell-Yan process for instance, the experimental
cross section attributed to the process is about a factor of 2 to 3
greater than what one predicts by using the lowest-order Feynman diagram.
A phenomenological $K$-factor is introduced by which the lowest-order
results must be multiplied in order to bring the lowest-order QCD
predictions into agreement with experiment.  Higher-order QCD
corrections to the Drell-Yan cross section, up to $O(\alpha_{\rm s})$
\cite{Kub80} and $O(\alpha_{\rm s}^2)$
\cite{Ham91}, have been worked
out.  These investigations show that in the high energy limit of
massless quarks, the $K$ factor can be well accounted for by including
QCD corrections \cite{Kub80,Ham91}. The most important contribution to
the dilepton cross section due to the $\alpha_{\rm s}$-order (the
next-to-leading-order) diagrams is the vertex correction at the $q\bar
q\gamma^*$
vertex involving the initial-state interaction between $q$ and $\bar
q$ and the gluon radiation from $q$ and $\bar q$. According to
\cite{Fie89,Bro93,Gro86}, the vertex correction
leads to a correction factor equal to $(1+2 \pi \alpha_{\rm s}/3)$.
For a strong interaction coupling constant  $\alpha_{\rm s}=0.3$, this vertex
correction gives a factor of about 1.7.  Additional contributions from
the $\alpha_{\rm s}$-order Compton diagrams bring the Drell-Yan
$K$-factor within the observed range of 1.6 to 2.8.  For charm and
heavy-quark production, the cross section for the lowest-order QCD was
given in \cite{Glu78,Com79}. Higher-order QCD corrections to the cross
sections and single-particle inclusive differential cross sections
have also been obtained \cite{Nas89a,Nas89b,Bee91,Col91}.  If one
includes the next-to-leading-order of QCD, the $K$-factor is found to
range from 2 to 3, depending on which heavy flavor is produced.  The
threshold behavior for heavy-quark production due to the distortion of
the $Q$-$\bar Q$ relative wave function has been examined in Refs.
\cite{App75,Bar80,Gus88,Fad88,Fad90}.  The correction is quite
large near the threshold.

 We focus in this work on basic reaction processes involving $q$ and
$\bar q$ in the initial and final states.  Other reactions involving
gluons are important part of the dynamics in the plasma or partons
and will be the subject of our subsequent studies.  We
investigate here the corrections to the tree-level cross sections
involving the  production and annihilation of $q \bar q$ at all relative
energies.  For low energies, we seek an analytical vertex correction
factor arising from wave function distortion by virtue of the
$q$-$\bar q$ color potential.  The distortion correction depends on
$\alpha_s/v$ where $v$ is the magnitude of the asymptotic relative
velocity.  It is the most important correction to the tree-level
descriptions at energies where masses cannot be ignored, as has been
earlier investigated for heavy quark production processes
\cite{App75,Bar80,Gus88,Fad88,Fad90}.  At high energies, we use the
well-known PQCD results which include the initial- or final-state
interactions in addition to real gluon radiations.  We shall seek an
interpolation to join the vertex correction factor from low energies
to these well-known perturbative QCD vertex correction factors at high
energies.  For the purpose of interpolation, we shall include the
proper relativistic kinematics and follow the interpolation procedure
suggested by Schwinger \cite{Sch73}.  We
obtain a simple analytical semi-empirical
correction factor which can be used over
the whole range of energies for both the annihilation and the
production of a quark-antiquark pair.

This paper is organized as follows.  In Section II,
we first give a general discussion on
the correction factor at low relative velocities
arising from the distortion of the wave function
at the point of $q\bar q$ annihilation or production.  This correction
factor is found to be an enhancement factor for the color-singlet
states
and a suppression factor for the color-octet states.
The interpolation of this result with the known perturbative QCD
correction factor at high energies provides  correction  factors which
can be applied over the whole
range of relative velocities.  In Section 3, we show how we can use
 vertex correction factors to improve the lowest-order
Drell-Yan cross
sections.  We discuss the use of these correction factors for dilepton
production in the quark-gluon plasma
in Section 4.  The correction factors lead to an enhancement of
the dilepton production probability in the plasma.
In Section  5, we examine the
 $q\bar q \rightarrow Q \bar Q$ process, where both $q\bar q$
annihilation
and $Q \bar Q$ production
are found to be
associated  with suppressive
correction factors.
Section 6 summarizes the present discussions.

\section{ General Consideration of the Corrective  Factor for
${ {\lowercase{ q }}} \bar  {\lowercase{ q }}$ Annihilation and Production}

What is the effective potential between the quark and the antiquark for
continuum states that populate  the production and annihilation
vertices?
The processes of production and annihilation are
characterized by a region of very small relative $q$-$\bar q$
separations,
at a linear distance of $\sim \alpha/\sqrt{s}$ for a virtual
photon intermediate state
and a distance of $\sim \alpha_s/\sqrt{s}$ for a virtual
gluon intermediate state.
The effective part of the $q$-$\bar q$ interaction in this region of
small
relative separation associated with
the annihilation or production process
is
the inverse-$r$, Coulomb-like term. The
linear part of the potential, that serves to effect the binding and impose
confinement, controls the large $r$ behavior and needs not be
considered
in the
collapsed spatial zone comprising the annihilation or production vertex.
Therefore, we consider the interaction between a quark and  an
antiquark with an invariant mass $Q=\sqrt{s}$ as described by
an effective Coulomb-type potential  $A=(A_0,\bbox{0})$ involving
their relative coordinate $r$
\cite{App75,Bar80,Gus88,Fad88,Fad90}
\begin{eqnarray}
\label{eq:1}
A_0(r)=-{ \alpha_{\rm eff} \over r},
\end{eqnarray}
where $\alpha_{\rm eff}$ is the
effective strong-interaction coupling constant,
related to the strong interaction coupling constant $\alpha_{\rm s}$ by
the color factor $C_f$,
\begin{eqnarray}
\alpha_{\rm eff}={C_f\alpha_{\rm s} } \,.
\end{eqnarray}
The running  coupling constant is  \cite{Fie89,Bro93}
\begin{eqnarray}
\alpha_{\rm s} = { 12 \pi \over (33 - 2 n_f ) \ln (Q^2/\Lambda^2) }\,,
\end{eqnarray}
where
the flavor number $n_f$ can be parametrized  to be
$$n_f = \sum_{q=u,d,s,c,b,t}  \biggl ( 1 - {4 m_q^2 \over Q^2} \biggr ) ^{1/2}
\theta  ( 1 - {4 m_q^2 \over Q^2} ) \,,
$$
with the step function $\theta$ and
 the threshold behavior inferred from that of  $q \bar q \rightarrow Q
{\bar Q}$ cross section.

When $q$ and $\bar q$ are in their
relative color-singlet states, the interaction  is attractive and
the color factor $C_f$ is $4/3$ \cite{Fie89,Won94}.
When $q$ and $\bar q$ are in their
relative color-octet  states, the interaction  is repulsive  and
the color factor $C_f$ is $-1/6$ \cite{Fie89}.

It is worth mentioning that the notion of a potential is a useful
concept, as one can infer from the success of the quasi-potential of
Tordorov \cite{Tor71} and Crater and van Alstine's
relativistic constraint dynamics
to describe  $q \bar q$ and $e^+e^-$
as a relativistic two-body problem \cite{Crax,Cra92}.
In these descriptions of constraint dynamics, the compatibility of the
two-body equations of motion requires that
 the two-body potential must be expressed in terms of the relative
four-coordinate $x_{\perp}$, orthogonal to the total four-momentum
$P$.  It is therefore convenient to work in the center-of-mass
coordinate system, where $P=(\sqrt{s}, {\bbox 0})$
and the four-coordinate
$x_{\perp}$ is the
relative coordinate $r$ used in  Eq.\ (\ref{eq:1}).
In this coordinate system, the two-body system is described by an
energy for relative motion
given by
\begin{eqnarray}
\label{eq:p0}
p^0=\epsilon_\omega= { s - 2 m_q^2 \over 2 \sqrt{s} }\,,
\end{eqnarray}
and the relativistic generalization of
the reduced mass of the system is $m_\omega=m_q^2/\sqrt{s}$.
The corresponding value of the on-shell relative
momentum squared at $r\rightarrow \infty$ is
\begin{eqnarray}
\label{eq:b2}
\bbox{p}^2=b^2(s)=\epsilon_w^2 - m_\omega^2={ s^2-4sm_q \over 4s} \,.
\end{eqnarray}
(See Eqs. (2.13a), (2.13b), and (2.13c) of Ref.\
\cite{Cra92}).

A proper treatment of the relativistic two-body problem will require
the solution of the two-body Dirac equation as formulated in Ref.\
\cite{Crax,Cra92} which is rather complicated.
To gain a simple insight into the effect of
the interaction between the quark and the antiquark,
we shall treat the problem approximately as a
single fermion spinor  in the  external field of Eq.\ (\ref{eq:1})
for which the solution is already known \cite{Akh65},
with the stipulation of the correct relativistic
kinematics to
describe
the relative motion between the quark and the
antiquark.
As is well-known, upon using the usual approximation as given by Ref.
\cite{Akh65},
the wave function which  satisfies the  correct boundary
condition of an incident plane wave is
\begin{eqnarray}
\label{eq:60}
\psi = N e^{i {\bf p} \cdot {\bf r}}
(1- { i \over  2 \epsilon_w} \bbox{\alpha} \cdot \nabla )
u F( i \xi, 1 , i( | \bbox{p}|r - \bbox{p}\cdot \bbox{r})),
\end{eqnarray}
where $u$ is the spinor for a  quark, $F$ is the confluent hypergeometrical
function, $N$ is
the normalization constant given by
\begin{eqnarray}
|N|^2=
{ 2 \pi \xi \over 1- e^{-2 \pi \xi}  },
\end{eqnarray}
and the parameter $\xi$ (which can be positive or negative)
is related to the coupling constant $\alpha_{\rm eff}$
and the magnitude of the asymptotic relative velocity $v$ by
\begin{eqnarray}
\label{eq:xi}
\xi={ \alpha_{\rm eff} \over v }.
\end{eqnarray}
The magnitude of the asymptotic
relative velocity $v$ is the ratio of the  momentum $b$ to
the energy $\epsilon_w$ in the system of relative coordinates.
Following Todorov \cite{Tor71} and Eqs.\ (21.13a)-(2.13c) of
Crater $et~al$  \cite{Cra92}, the
relative velocity for the quark and the antiquark
in their center-of-mass system obtained by using Eqs.\ (\ref{eq:p0})
and (\ref{eq:b2}) is
\begin{eqnarray}
\label{eq:65}
v= {(s^2 - 4s m_q^2)^{1/2} \over s - 2 m_q^2 },
\end{eqnarray}
which gives $v \sim 2\sqrt{1-4m_q^2/s}$ when $\sqrt{s} \sim 2m_q$
and $v\rightarrow 1$ when $s \rightarrow \infty$.
The quantity $v$ is the asymptotic relative velocity obtained with the
proper relativistic kinematics and differs from the quantity
$\beta=\sqrt{1-4 m_q^2 /s}$ defined by Fadin $et~al.$ \cite{Fad90}.
For the Drell-Yan $q\bar q$  annihilation processes or for
the heavy-quark pair production
processes, a quark and an antiquark are first annihilated into (or
produced from) a virtual photon
or a virtual gluon.  As annihilation (or production) takes place in the
vicinity of the
region around $r=0$, the probability for the annihilation process is
proportional to the absolute square of the wave function at $r=0$. When one
averages over the spins of the quark, the square of the wave function
at the origin in the case with interaction $A_0$ is
\begin{eqnarray}
|\psi(0)|_{A_0}^2= N^2 \biggl [ 1 +
\biggl (  { \alpha_{\rm eff} \over E v} \biggr )^2 p^2 {(1- \cos
\theta) \over 2} \biggr ] ,
\end{eqnarray}
where $\theta$ is the angle between the incident momentum $\bbox{p}$
and $\bbox{r}$.
After averaging over $\theta$,
 we have
\begin{eqnarray}
|\psi(0)|_{A_0}^2=
{ 2 \pi \xi \over 1- e^{-2 \pi \xi}  }
( 1 + {\alpha_{\rm eff}^2 \over 2}).
\end{eqnarray}
The above result was  obtained by  treating the $q \bar q$ system
approximately
as a fermion spinor in the field (\ref{eq:1}) of a spinless particle
and the  correction
term
$\alpha_{\rm eff}^2/2$ inside the bracket
arises from the spin  of one of
the two particles. When  the
spin  of the other particle is
taken into account, there is an
additional $\alpha_{\rm eff}^2/2$ contribution,
 and the square of the wave function at the origin is modified to
\begin{eqnarray}
\label{eq:9}
|\psi(0)|_{A_0}^2=
{ 2 \pi \xi \over 1- e^{-2 \pi \xi}  }
( 1 + \alpha_{\rm eff}^2 ).
\end{eqnarray}
On the other hand, in the perturbative expansion
in  which there is no interaction between the quark and the
antiquark in the lowest-order approximation, the square of the wave function
$|\psi(0)|_o^2$ is unity.  Therefore, by considering the
interaction between the quark and the antiquark as arising from an effective
Coulomb-like interaction,
the annihilation or production cross
section
is modified  by the factor, which is a generalization of the
`Gamow-Sommerfeld' factor \cite{Gam28,Som39}, given by
\begin{eqnarray}
\label{eq:80}
K={ |\psi(0)|_{A_0}^2 \over
            |\psi(0)|_{o}^2 }
=
{ 2 \pi \xi \over 1- e^{-2 \pi \xi}  }
( 1 + \alpha_{\rm eff}^2 ),
\end{eqnarray}

The expansion of the
above  $K$ factor arising from the distorted wave function
gives
\begin{eqnarray}
\label{eq:15}
K= \biggl [1 + {2\pi\xi \over 2} +{(2 \pi  \xi)^2 \over 12}
- { (2 \pi \xi)^4 \over 720}
+{(2 \pi \xi)^6 \over 30240} - { (2 \pi \xi)^8 \over 1209600}
+... \biggr ](1+  \alpha_{\rm eff}^2 )\,,
\end{eqnarray}
which contains higher order terms with alternating signs.
It is clear that for the case when
$|\pi C_f \alpha_s/v |$ is much greater than 1, the convergence of the
 perturbation expansion as a power of $\alpha_s$
will be very slow.  Many higher-order terms
are needed to approach  the full  result  Eq.\ (\ref{eq:80}).
For that case,  a perturbative treatment of the correction factor is
not useful and a non-perturbative treatment such as
presented here in terms of wave-function distortion represented by
Eq.\ (\ref{eq:80}) is  a
meaningful concept.

The correction factor  Eq.\ (\ref{eq:80}) has been obtained
by considering only
the exchange of virtual
gluons
between the quark and the antiquark.
The emission of real soft gluons is not taken into account.
Because
radiative corrections involving the emission of real gluons
is unimportant
for $q\bar q$ systems with small
relative velocities,
 the results of Eq. (\ref{eq:80}), without the inclusion of real
soft gluons, is applicable to the region of low relative velocities.

At very-high relative velocities, besides the distortion effect arising
from the initial-state and final-state interactions
between the quark and the antiquark, it is necessary to include
the
radiation of real gluons.
In that case,
the amplitude from
the emission of soft  gluons  interferes
with the amplitude arising from the   exchange of a
virtual gluon
 between
the quark and the antiquark.
At very high relative velocities for which the masses of the quarks
can be neglected,
perturbative QCD
calculations
have been carried out by including the exchange of a virtual gluon
between the quark and the antiquark and
the emission of real gluons from the two particles
 \cite{Kub80,Ham91,Fie89,Bro93}.
These high-energy  vertex correction $K$-factors are different for
$q\bar q$ annihilation
or for $q\bar q$ production.
As the vertex correction gives the dominant contribution,
we shall consider the approximation where the phenomenological
correction factor $K$ is
given just by the vertex correction.
Up to the first order in $\alpha_s$
in the high-energy limit
 of massless quarks,
the vertex $K$-factor
for $q\bar q$ annihilation
is
 \cite{Kub80}
\begin{eqnarray}
\label{eq:in}
K &&=1+ \pi   \alpha_{\rm eff} \biggl ( {1 \over 2 \pi^2}
+ { 5 \over 6} \biggr )
\nonumber\\
&&=1+ \pi  \alpha_{\rm eff}
{}~  \times 0.884,
\end{eqnarray}
which is about 2 for a typical $\alpha_{\rm eff}=0.4$ for
color-singlet states.
The vertex correction $K$-factor
for $q\bar q$ production  in the $e^+e^- \rightarrow q \bar q$
process
is \cite{Fie89,Bro93}
\begin{eqnarray}
K =1+   {  \alpha_{\rm s} \over  \pi}
\,,
\end{eqnarray}
which is about 1.1 for $\alpha_{\rm eff}=0.4$.
On account of the fact that the singlet color factor for
$e^+e^-$ annihilation is $4/3$, the correction factor for
$q\bar q$ production is
\begin{eqnarray}
\label{eq:out}
K =
1+   { 3 \alpha_{\rm eff} \over 4 \pi} \,.
\end{eqnarray}
Thus, for $q\bar q$ in a color-singlet state at high energies, the high-order
QCD
corrections are large for
$q\bar q$ annihilation and small for
$q\bar q$ production.
In contrast, at low relative velocities, Eq.\ (\ref{eq:80})
gives equally large QCD  corrections for
$q\bar q$ annihilation or production.

Knowing the two limits of the correction factors from Eqs.\
(\ref{eq:80}), (\ref{eq:in}),
and (\ref{eq:out}), we can follow the interpolation procedure
suggested
 by Schwinger
\cite{Sch73}
and used by Barnett $et~al.$ \cite{Bar80}.
We can  obtain a semi-empirical correction factor which can be used for
the whole range of relative velocities.
For $q\bar q$ annihilation in the initial state,
we define a function $f^{(i)}(v)$ by
\begin{eqnarray}
f^{(i)}(v)= \alpha_{\rm eff} \biggl [  { 1  \over v }
+ v  \biggl (  - 1 +
{ 1 \over 2 \pi^2}
+ { 5  \over 6} \biggr ) \biggr ] \,.
\end{eqnarray}
For $q\bar q$ production in the final state,
we define a function $f^{(f)}(v)$ by
\begin{eqnarray}
\label{eq:130}
f^{(f)}(v)= \alpha_{\rm eff} \biggl [  { 1 \over v }
+ v
 \biggl ( - 1 + {3 \over 4 \pi^2}
\biggr ) \biggr ]  \,.
\end{eqnarray}
Associated with the annihilation or production of a $q\bar q$ pair at a
center-of-mass energy
$\sqrt{s}$,
we propose a semi-empirical vertex $K$-factor
as
\begin{eqnarray}
\label{eq:150}
K^{(i,f)}(q)
={ 2 \pi f^{(i,f)}(v) \over 1- \exp \{ {-2 \pi f^{(i,f)}(v)} \}  }
( 1+ \alpha_{\rm eff}^2 )  \,,
\end{eqnarray}
where the flavor label  $q$ in $K^{(i)}(q)$ is
included  to indicate that $K^{(i)}$
depends on the quark mass $m_q$,
the superscript $(i)$ and $(f)$ denotes
$q\bar q$ $\underline i$nitial-state annihilation
and $\underline f$inal-state production
 respectively, and the magnitude of the relative velocity
$v$ is given by Eq.\
(\ref{eq:65}).
This vertex correction factor agrees with the results of Eq.\
(\ref{eq:80})
for low relative velocities.  For high relative velocities and
$|\pi \alpha_{eff}| << 1$, Eq.\ (\ref{eq:150}) gives
\begin{eqnarray}
K^{(i,f)} \rightarrow 1 + \pi f^{(i,f)}\,,
\end{eqnarray}
and Eq.\ (\ref{eq:150}) agrees with
the well-known results
of Eqs.\ (\ref{eq:in}) and (\ref{eq:out})
from perturbative QCD for high relative velocities,
up to the first order in $\alpha_s$.
Thus,  Eq.\ (\ref{eq:150}) can provide a good description of
 the vertex corrections for  $q\bar q$ systems over the whole  range
of relative velocities, for
$q\bar q$ annihilation or $q\bar q$ production.
Because the vertex correction gives the dominant QCD correction, the
vertex correction factor (\ref{eq:150})
can be used to give the corrections to  lowest-order QCD results when
a $q\bar q$ pair is annihilated or produced. It
could equally well be used with a subscript `v'
to identify its association
with the vertex correction and distinguish it from the usual phenomenological
$K$ factor that usually quantifies the QCD corrections in the massless limit
only. However, it has been retained as $K$ for compactness and to allow easy
applicability of the correction.

\section{ Drell-Yan Processes }

As an example of the application of
 the correction factor $K^{(i,f)}$ of Eq.\ (\ref{eq:150}),
we can use it to improve theoretical
tree-level estimates of the Drell-Yan cross section.
In the Drell-Yan process, a $q\bar q$ pair is annihilated in the
initial state,
and the appropriate correction factor is $K^{(i)}$.
As the virtual photon
in the  $q\bar q\gamma^*$ vertex selects the color-singlet
combination of the annihilating $q\bar q$ pair, the corresponding
color factor
 $C_f$ is 4/3, and the effective coupling constant needed to
calculate
$K^{(i)}$ in Eq.\ (13) is $\alpha_{\rm eff}=4 \alpha_s/3$.

If only the lowest theoretical estimate $d\sigma^{\rm LO}/dQ^2 dy$
is available, an improved determination  of the Drell-Yan cross section for $q
\bar q \rightarrow l^+ l^-$ can be
provided by
$${d\sigma \over dQ^2 dy }= K^{(i)}(q)
{ d\sigma^{\rm LO} \over dQ^2 dy }.$$
In Fig.\ 1, we show  $K^{(i)}(q)$
as a function of the invariant mass $Q=\sqrt{s}$ of the
dilepton pair calculated with $\Lambda=0.3$ GeV.
For a $q\bar q$ pair in a color-singlet state at
high relative velocities, the $K^{(i)}$ factor is about
$2.2$ which  agrees with the experimental value of 1.6 to 2.8
\cite{Bro93}
in the Drell-Yan process.  For  color-singlet states with low relative
velocities, the correction factor is quite large.
The correction factor for a $q\bar q$ pair in the color-octet state
approaches 0.9 as $\sqrt{s}$
increases and it deviates from this constant behavior at low relative
velocities when the energy is near the $q\bar q$ mass threshold.

When the next-to-leading order results, including other diagrams
such as Compton diagrams  are available, the estimate of the
Drell-Yan cross
section can be improved to include the distortion corrections as well.
By  following a  procedure which was suggested by Harris and Brown
\cite{Har57},
 we can avoid  double
counting.

Accordingly, the next-to-leading order results of Kubar
$et~al.$ for the Drell-Yan cross section can be  modified to include distortion
effects by using the $K$-factor as given by Eq.\ (\ref{eq:80}) prior  to
interpolation with the high energy QCD corrections.
Using most of the notation of Kubar $et~al.$, the cross section for the
production of an $l^+l^-$ pair with an invariant mass squared $Q^2$
and a rapidity $y$ for the collision of two hadrons is
\begin{eqnarray}
{ d \sigma \over dQ^2 dy}
=
{ d \sigma^{\rm LO} \over dQ^2 dy}
+{ d \sigma^{\rm A}  \over dQ^2 dy}
+{ d \sigma^{\rm C}  \over dQ^2 dy} \,,
\end{eqnarray}
with
\begin{eqnarray}
{ d \sigma^{i} \over dQ^2 dy}
= \int dt_1 dt_2 \sum_{f}
{ d {\hat{ \sigma}}^{i} \over  dQ^2 dy}  Q_f(t_1,t_2),
\end{eqnarray}
where $i={\rm LO,~A,~C}$ represent the contributions from the lowest
order Drell-Yan diagram, the annihilation diagrams, and the Compton
diagrams,
respectively.  The function $Q_f$ is the product of the quark and
antiquark structure functions of  the colliding hadrons,
\begin{eqnarray}
Q_f(t_1,t_2)=
 q_{10}^f(t_1) {\bar q}_{20}^f(t_2)
+{\bar q}_{10}^f(t_1) { q}_{20}^f(t_2)\,,
\end{eqnarray}
and $f$ is the flavor label.

When the distortion
due
to the wave function is taken into account, the result of Kubar $et~al$
\cite{Kub80} can be
modified to be
\begin{eqnarray}
{ d {\hat \sigma}^{\rm LO} \over dQ^2 dy}
&&+{ d {\hat \sigma}^{\rm A}  \over dQ^2 dy}
\nonumber\\
&&= { 4 \pi \alpha^2 \over 9 Q^2 s}
\biggl ( { e _f \over e } \biggr ) ^2
\biggl [
\delta(t_1-x_1) \delta(t_2-x_2 ) K^{(i)}
\nonumber\\
&&~~~~~~~~~~~~~~~~~~~~\times
 \biggl \{ 1
+  { |\alpha_{\rm eff}| \over 2 \pi }
[ - { 3 \over 2} \ln { x_1 x_2 \over (1-x_1)
(1-x_2) } + 2 \ln { x_1 \over 1-x_1} \ln { x_2 \over 1-x_2 } ]  \biggr \}
\nonumber\\
&&+ { |\alpha_{\rm eff}| \over 2 \pi} \delta(t_2-x_2)
\biggl ( { t_1^2 + x_1^2  \over t_1^2( t_1-x_1)_+ }
+ \ln { 2 x_1 (1-x_2) \over x_2 ( t_1- x_1) }
+ { 3 \over 2 (t_1-x_1)_+}  - { 2 \over t_1} - {3 x_1 \over t_1^2}
\biggr )
+ ( 1 \leftrightarrow 2 ) \nonumber\\
&&
+ { |\alpha_{\rm eff}| \over  \pi}
\biggl ( { G^{\rm A} (t_1,t_2) \over [(t_1-x_1) ( t_2-x_2)]_+}
+ H^{\rm A} (t_1, t_2) \biggr ) \biggr ]\,,
\end{eqnarray}
where $K^{(i)}$ is now given by Eq.\ (\ref{eq:80}),  the distributions
$1/(t-x)_+$, and $1/[(t_1-x_1)(t_2-x_2)]_+$ are
defined by
$$
\int_x^1 dt { f(t) \over (t-x)_+ }
= \int_x^1 dt { f(t)- f(x) \over t-x} \,,
$$
and
$$
\int_{x_1}^1 dt_1 \int_{x_2}^1 dt_2
{ f(t_1,t_2) \over [(t_1-x_1)(t_2-x_2)]_+}
=
\int_{x_1}^1 dt_1 \int_{x_2}^1 dt_2
{ f(t_1,t_2) - f(t_1,x_2) - f(x_1, t_2) + f (x_1, x_2)
\over (t_1-x_1) ( t_2 - x_2)} \,.
$$

In another perturbative treatment of the Drell-Yan process,
an estimate  of the first-order vertex correction
leads to a vertex $K$ factor of the form
 \cite{Fie89,Bro93,Gro86}
\begin{eqnarray}
\label{eq:kp}
K'=1 + 2 \pi \alpha_{\rm
s} / 3\,,
\end{eqnarray}
and the generalization to higher-orders of the form
\begin{eqnarray}
\label{eq:17}
K'= e^{ 2 \pi \alpha_{\rm s} / 3}\,.
\end{eqnarray}
However, this result has been obtained only
in the high energy limit and
may not be applied to the cases of low dilepton
invariant masses.  Furthermore, the vertex correction factor of Eq.\
(\ref{eq:kp})
appears to be only a part of the whole vertex correction factor of Eq.\
(\ref{eq:in}).

Combined with the running coupling constant, the use of the correction
factor
Eq.\ (\ref{eq:150})
allows us to probe lower dilepton invariant masses and the important
low-$x$ region with better justification than massless limits.

\section{Thermal Dilepton Production}

     The importance of the possibility of hadron  matter going through a phase
transition into a
deconfined quark-gluon plasma state, during high-energy heavy-ion collisions
has been highlighted earlier \cite{Won94}.
While the basic process
leading to the formation of dileptons is the same as that for the
Drell-Yan
process,
the distributions and the characteristics of the annihilating quarks are quite
different for the two cases. The final state dileptons are expected to carry
information on the environment of the annihilating quark-antiquark pair and
therefore serve as a probe of the thermodynamical state of the quark-gluon
plasma.

     The rate for the production of dileptons with an invariant mass
 $Q$ per unit four-volume in a thermalized quark-gluon plasma
is sensitive to the temperature $T$ of the system and can be written in the
form
\cite{Kaj86,Won94}
\begin{eqnarray}
\label{eq:14}
{dN_{l^+l^-} \over dQ^2  d^4x} \sim N_c N_s^2
\sum_{q=u,d,s...} \biggl ({e_{q} \over e} \biggr )^2
{\sigma_q (Q) \over 2(2\pi)^4} Q^2
 \sqrt{ 1-{4m_q^2 \over Q^2} }
 T Q K_1\biggl ({Q\over T} \biggr ) \,,
\end{eqnarray}
where
$\sigma_q(Q)$ is the lowest order
$q\bar q \rightarrow l^+ l^-$ cross section at the center-of-mass
energy $Q$ given by \cite{Won94}
\begin{eqnarray}
\label{eq:50}
\sigma_q(Q)
=
{4 \pi \over 3} { \alpha^2 \over Q^2}
\biggl (1 - {4 m_q^2 \over Q^2} \biggr )^{\!\!-{1\over2}}
\!\!\sqrt{ 1-{4m_l^2 \over Q^2}}
\biggl ( 1+ 2{ m_q^2+m_l^2 \over Q^2} +4{m_q^2 m_l^2 \over Q^4} \biggr )
\,,
\end{eqnarray}
$m_l,~ m_q$ are  the rest masses  of the lepton $l$
and
the quark  $q$ respectively, and
$K_1$ is the modified Bessel function of first order.

       In order to be meaningful experimental signals for the detection of QGP
states, the thermal dileptons must be clearly delineated from other sources of
dileptons, particularly the Drell-Yan background. It becomes important
therefore to have good estimates of the dilepton production from the different
sources at matching levels of accuracy. It is hence necessary to modify
equation (\ref{eq:14}), to incorporate the higher-order
QCD corrections into the rate for
dilepton production in the quark-gluon plasma.

The interpolation from low to high relative velocities extends the
usefulness of the correction factors Eq.\ ({\ref{eq:150}).  Therefore,
we may introduce the same correction factor $K^{(i)}$ as used in the
DY case to account for high order QCD effects.  The modified rate of
dilepton production becomes
\begin{eqnarray}
{dN_{l^+l^-} \over dQ^2  d^4x} \sim N_c N_s^2
\sum_{q=u,d,s...} \biggl ({e_{q} \over e} \biggr )^2 {K^{(i)}(q )
\sigma_q(Q) \over 2(2\pi)^4} Q^2
\biggl
 (1-{4m_q^2 \over Q^2} \biggr )^{1 \over 2}
TQ K_1\biggl ({Q\over T} \biggr ) \,.
\end{eqnarray}
The virtual photon annihilation mode
selects  the annihilating $q\bar q$ pair
to be in the  color-singlet states and thus
the interaction is attractive.
The distortion effect leads to an enhancement factor $K^{(i)}$
as shown in Fig.\ 1.
Because the coupling constant increases and the relative velocity
decreases as the invariant mass decreases, the correction factor rises
considerably with the decrease of invariant mass.  The effect of the
higher order QCD corrections
is to enhance the tree-level dilepton cross section
by a factor of about 4 at $\sqrt{s}=1$ GeV, and by a factor of about 3
at $\sqrt{s}=2-3$ GeV.
Higher order QCD corrections
will increase the strength of the dilepton signal and may serve to
enhance the prospects for its detection in ongoing and planned
experiments \cite{Phe93}.

       It may be remarked that in the literature, a correction $K$
factor has sometimes been used for thermal production  as
$[ 1 + (\alpha_{\rm s}/ \pi)(1+aT^2/Q^2)]$ (Eq.\ (5.1) of Ref.\ \cite{Alt89}),
which
coincides approximately with $K^{(f)}$  that includes
 the first-order correction for the
reverse $l^+\l^-\rightarrow q\bar q $ process \cite{Fie89}, but is
smaller than the correction factor $K^{(i)}$.  Recent
calculations, for the first order correction, using the thermal mass,
predict a much larger  $K$ factor similar to our $K^{(i)}$
in magnitude
\cite{Alt92}.

   The question of screening in the quark gluon plasma requires a
comment at this stage.
In the plasma, the color charge of the constituents of the plasma is
subject to Debye screening which is characterized by the Debye
screening length $\lambda_D$ depending on the temperature $T$.  For
the quark-gluon plasma with a flavor number $N_f$ and $N_c=3$, the
lowest-order perturbative QCD theory gives \cite{Gro81}
\begin{eqnarray}
\lambda_D({\rm PQCD})
=
{1 \over \sqrt{  \bigl ( {  N_c \over 3}+ {N_f \over 6 }\bigr ) g^2  }
{}~T} \,.
\end{eqnarray}
For a coupling constant
$\alpha_s=g^2/4\pi=0.5$ and $N_f=3$, the Debye screening
length at a temperature of 200 MeV is about $\lambda_D \sim 0.4$ fm.
On the other hand, the electromagnetic annihilation into
dileptons occurs within an extremely collapsed space zone
characterized by the linear $q$-$\bar q$ distance
$\sim \alpha/\sqrt{s}$
which should be compared with the
the Debye screening
radius $\lambda_D$ in the plasma.
As the
annihilation distance $\alpha/\sqrt{s}$
is about $0.0029$ fm for the annihilation of a light
quark-antiquark pair at 0.5 GeV, and is much smaller than the Debye
screening length $\lambda_D$, the interaction between the quark and
the antiquark is not much affected by Debye screening in the region
where annihilation occurs.  Therefore, we expect that our correction
factor for dilepton production by $q \bar q$ annihilation in the
plasma will not be modified much by the addition of Debye screening
corrections.

\section{ Heavy-Quark Production}

       Heavy quark production in nuclear collisions, whether routed through
intermediate QGP states or not, occurs either by $q\bar q$
annihilation  ($q\bar q \rightarrow Q\bar Q$) or by gluon-gluon fusion,
($g g \rightarrow Q\bar Q$). In the first case, the final state heavy
quark-antiquark pair are required to be in a color-octet outgoing state, due to
the
color-octet nature of the intermediate gluon states, while in the latter, they
may emerge in either the color-singlet or color-octet states. We shall limit
our
investigation in this paper to the former process, and discuss the
gluon-gluon
process in a future  publication.

    The cross section for $q\bar q \rightarrow Q\bar Q$ can be written at
tree-level as \cite{Won94}
\begin{eqnarray}
\label{eq:250}
\sigma(s)
=
{ 8 \pi \alpha_s^2 \over 81 s}
\biggl (1 - {4 m_q^2 \over s} \biggr )^{\!\!-{1\over2}}
\!\!\sqrt{ 1-{4m_Q^2 \over s}}
\biggl ( 1+ 2{ m_q^2+m_Q^2 \over s} +4{m_q^2 m_Q^2 \over s^2} \biggr )
\,.
\end{eqnarray}
   In the  $q\bar q \rightarrow g^* \rightarrow Q\bar Q$ process,
there are two
strong-interaction vertices,
$q\bar q  g^*$ and $g^*  Q\bar Q$.
Both of these require  QCD
corrections, along with the other radiative corrections involving the external
quark lines and their connections with the virtual gluon propagator.

     In the present work, we attempt to
 correct for the two interaction vertices simultaneously, to include
the wave-function distortion effect at low relative velocities  and additional
real gluon radiation at high relative velocities, in the same way as in
our earlier corrections for the DY process.

   The effective interaction between the quark and the
antiquark can be written for each vertex as in section II, and we  can
similarly write a $K$ factor for each vertex. The first vertex,
$q\bar q  g^*$, occurs with the annihilation of a $q\bar q$ pair in
the initial state
and is associated with the
$K^{(i)}(q)$ correction factor.  The vertex
is identical to the DY case except for the replacement of the virtual photon
 by the virtual gluon.
Because of the color-octet nature of the intermediate gluon,
the quark-antiquark pair
must likewise be in the color-octet state.
The correction factor
$K^{(i)}(q)$ for the annihilation of this  color-octet
state corresponds to  an effective repulsive potential
with the  color factor
 $C_{ f}$ equal to   $-1/6$.  As we note from Fig.\ 1, it has the
value of about 0.9 at high energies.

   The  second vertex,
$g^*  Q\bar Q$, occurs with the production
of the $Q\bar Q$ pair in the final state
and is associated
with
the $K^{(f)}(Q)$ correction factor
given by Eq.\
(\ref{eq:150}),
 having the appropriate function $f^{(f)}(v)$ of Eq.\ (\ref{eq:150}).
Since the color factor is same as the
first vertex, this correction is also suppressive.

   Corrected for wave function distortion at both vertices, the cross
section (\ref{eq:250}) is amended to
\begin{eqnarray}
\sigma ( s)=
K^{(i)}(q)  K^{(f)}(Q)
{8 \pi  \alpha_s^2 \over 81 s}
\biggl (1 - {4 m_q^2 \over s} \biggr )^{\!\!-{1\over2}}
\!\!\sqrt{ 1-{4m_Q^2 \over s}}
\biggl ( 1+ 2{ m_q^2+m_Q^2 \over s} +4{m_q^2 m_Q^2 \over s^2} \biggr )
\,.
\end{eqnarray}
In the range of invariant mass close to the masses of the heavy quarks
in question, the ideal analysis envisages correct inclusion of the
heavy quark masses.  The vertex factor $K^{(f)}(Q)$ accordingly
acquires a sensitivity to the mass.  In Fig. 2, we display the
correction factor $K^{(f)}$ for the production of $q\bar q$ pairs of
various flavors in color-singlet or color-octet states, calculated
with $\Lambda=0.3$ GeV.  The values of the correction factors are
constants at relativistic relative velocities, but begin to depart
from the constant values as the invariant mass approaches the
threshold, corresponding to lower relative velocities between the
quark and the antiquark.  Near the threshold, the enhancement of the
color-singlet states due to distortion is much more substantial than
the corresponding suppression of the production of color-octet states.

How does the Debye screening affect the correction factor?
The annihilation or production cross section in the  $q\bar q \rightarrow
Q \bar Q$ process
involves the strong coupling constant and is associated with an
annihilation or production  length
of the order of $\alpha_s/\sqrt{s}$
($cf.$ Eq.\ (\ref{eq:250})) which
should be compared with the
Debye
screening length $\lambda_D$
of about 0.4 fm at $T=200$ MeV.
The effect of Debye screening is important
when $\alpha_s/\sqrt{s}>> \lambda_D$;
it is unimportant
when $\alpha_s/\sqrt{s}<< \lambda_D$.
Accordingly, for quark-antiquark production involving the annihilation or the
production of light
$q\bar q$ pairs through a virtual gluon intermediate state,
the effect of Debye screening is important
when $\sqrt{s} << 0.25 $ GeV, assuming a coupling constant of
$\alpha_s=0.5$.
It is unimportant
when $\sqrt{s} >> 0.25$  GeV.
Thus, the correction factors can be used for $q\bar q$ pairs with
energies $\sqrt{s}$ much greater than  0.25 GeV.

\section{ Conclusions and Discussions}

At low energies of relative motion,
the quark-antiquark relative wave function
becomes distorted
in the
field of the  color potential acting between a quark and its antiquark
partner
\cite{App75,Bar80,Gus88,Fad88,Fad90}.
The distortion is strong  in the vicinity of their zero
separation and leads to a significant modification
of the cross section for their
annihilation or production.
When the quark-antiquark pair is in its color-singlet state, the
 interaction is highly attractive, and the cross section is
much enhanced.
On the other hand, when the quark-antiquark pair is in its color-octet
state, the interaction is repulsive and the cross section is
suppressed.
This enhancement or suppression can be quantified in terms of a vertex
correction factor $K$.
The distortion effect does not depend on whether the
quark-antiquark pair is annihilated or produced.
At high energies of relative motion such that the quark masses can be
neglected, the  addition of  real soft gluon
emission leads to  well-known PQCD vertex correction factors
which are different for $q\bar q$ annihilation or production \cite{Fie89}.
Following the  interpolation procedure of Schwinger to connect the
vertex correction factor   for low relative velocities, (obtained
from wave function  distortion),
to  the vertex correction factor for high relative velocities,
(obtained from perturbative
 QCD), we can
interpolate the correction
factor to  make it applicable to the whole range of relative velocities.
The correction factor enables us to correct tree-level QCD
calculations,
when a $q\bar q$ pair is annihilated or produced.
Its  inclusion improves both our
understanding of the underlying physics and the  accuracy
of the reaction cross sections.

       We wish to emphasize here the usefulness of our unified
description
to correct the lowest-order results  in different
environments. As
our $K^{(i,f)}$ does  not dependent on the structure functions or
constituent distributions
in $q\bar q$ annihilation
or the mode of $q\bar q$ production, it does not have to
be recalculated for every individual environment. It is sufficient to ensure
that the process considered is described by basic diagrams involving
$q \bar q$ annihilation or production.

As a test of the reliability of
the correction factor Eq.\ (\ref{eq:150}),
we apply the correction factor for $q \bar
q$ production to study the ratio $R=\sigma(e^+ e^- \rightarrow {\rm
hadrons}) /\sigma( e^+ e^- \rightarrow \mu^+ \mu^-)$.
For heavy quark production in the $e^+ e^- \rightarrow \gamma^*
\rightarrow  Q \bar Q$
process, $Q$ and $\bar Q$ are produced in the color-singlet final state.
The tree-level cross section must be multiplied by $K^{(f)}$ of Fig.\
2 to take into account higher-order QCD corrections.  The results of
Fig.\ 2 shows a very large enhancement near the threshold of
heavy-quark production.  This large enhancement is indeed observed and
the results Fig.\ 2 give good agreement with experimental ratio $R=
\sigma(e^+e^- \rightarrow {\rm hadrons}) / \sigma(e^+e^-
\rightarrow
\mu^+\mu^-) $  \cite{Cha94}.  The good agreement lends support to the
usefulness of Eq.\ (\ref{eq:150}) for
its application to other reaction processes.

Accordingly, one can
use the correction factors to improve
the tree-level estimates  of  many
cross sections.  For example, in dilepton production via the Drell-Yan
process in
nucleon-nucleon collisions, the correction factor $K^{(i)}$ is about 2.2 from
Fig.\ 1 at high energies which agrees with experimental values of 1.6 to 2.8
\cite{Bro93}.
 In the quark-gluon plasma, dilepton production from the collision of
quarks and antiquarks in the plasma occurs through the reactions $u
\bar u \rightarrow l^+ l^-$ and $d \bar d \rightarrow l^+ l^-$.
Because of our interpolation using Schwinger's procedure, the
semi-empirical correction factors of Eq.\ (\ref{eq:150}) is applicable
to these processes.  The tree-level cross sections for $u \bar u
\rightarrow l^+ l^-$ and $d \bar d \rightarrow l^+ l^-$ should be
corrected by multiplying them with the correction factor $K^{(i)}$, to
take into account the initial-state $u$-$\bar u$ or $d$-$\bar d$ color
interactions and radiations.  The results of Fig.\ 1 indicates that
this multiplicative correction factor for dilepton production in the
quark-gluon plasma is significantly large.  It increases as the
dilepton energy decreases.
The action of
of the correction factor is to enhance the tree-level dilepton cross
section by a factor of about 4 at $\sqrt{s}=1$ GeV, and by a factor of
about 3 at $\sqrt{s}=2-3$ GeV.

In the
process of $q \bar q \rightarrow s \bar s$ in the quark-gluon plasma,
where $q$ is a light $u$ or $d$ quark, the tree-level cross section
should be multiplied by the correction factors $K^{(i)}(q\bar q)
K^{(f)}(s \bar s)$ to take into account the initial- and final-state
interactions and gluon radiations.  Because these reactions take place
when the quark and antiquark are in the color-octet state, the
interaction is repulsive, and the two correction factors are less than
unity.  The deviation from unity is great near the $s \bar s$
threshold.  Another application of the present investigation is the
use of the correction factor to provide a better estimate of the
production of top quarks by the $e^+e^-$ annihilation, which will be
the subject of a separate publication \cite{Cha94}.

In the plasma, the color charge of the constituents of the plasma is
subject to Debye screening.  We shall investigate
quantitatively how the correction factors may be modified by the Debye
screening.  One can get a qualitative understanding whether the Debye
screening may lead to a large modification on the correction factor of
Eq.\ ({\ref{eq:150}) by comparing the magnitudes of length scale, over
which annihilation or production takes place, with the Debye screening
length.  If the length scale for annihilation or production is much
greater than the Debye screening length, then the color charge of one
interacting particle is effectively shielded from the other particle,
when the quark and the antiquark come to the region of annihilation or
production.  In this
case, the conrrection factor of Eq.\ (\ref{eq:150}) calculated with no
Debye screening will be much modified when the Debye screening is
taken into account.  On the other hand, when the length scale for
annihilation is much smaller than the Debye screening length, as in
the case of dilepton production in the quark-gluon plasma, the effect
of Debye screening is small.  Accordingly, the Debye screening will
not significantly modify the correction factor of Eq.\ (\ref{eq:150})
for dilepton production, and for the reaction $q\bar q \rightarrow Q
\bar Q$ with $\sqrt{s} >>  0.25$ GeV.  The Debye screening is
important for the reaction $q\bar q \rightarrow Q \bar Q$ with
$\sqrt{s} <<  0.25$ GeV.  Therefore, it remains reasonable to
apply the present results of the correction factor without considering
Debye screening to a substantial domain of reactions in the
quark-gluon plasma.

An important part of the dynamics in the plasma or partons involves
the reaction with gluons.  The evaluation of the complete set of
reaction cross sections in the plasma or partons will require the
investigation of next-to-lowest order corrections to gluon reaction
processes, which will be the subject of our future studies.  Just as
for a $q\bar q$ pair, initial- and final-state gluon-gluon and
gluon-quark interactions are expected to lead to large corrections to
cross sections for reactions involving $gg$ or $gq$ pairs.  A $gg$
pair can form many different color multiplets and the interaction
between a gluon and another gluon is attractive when they are in their
color-singlet and color-octet states.  It has been argued that gluon
dynamics may be described as massive spin-1 fields with the mass
generated dynamically through strong gluon-binding forces
\cite{Cor83}.  Investigations on the gluon-gluon
and gluon-quark correction factors using a gluon exchange potentials
as in \cite{Cor83,Hou84} will be of great interest.

Future refinement of the present investigation can be carried out by
including the variation of the coupling constant in modifying the
spatial dependence of the color potential between a quark and its
antiquark partner in the plasma.  The correction factor we have
obtained is based on an approximate simplified description by treating
the two-body problem as a single-particle problem with the proper
relativistic kinematics.  A better two-body Dirac equation based on
Crater and van Alstine's constraint dynamics \cite{Crax} has already
been worked out and used to examine electromagnetic interactions
\cite{Cra92}.  Its application to the present problem involving a
quark and an antiquark in a Coloumb-like interaction will be of great
interest.  Furthermore, the solution of Eq.\ (6) is actually obtained
by neglecting the $O(\alpha_{\rm eff}^2/r^2)$ term in the second-order
Schr\"odinger eqaution derived from the single-particle Dirac equation
\cite{Akh65}.  It will  be useful to include this term in future
investigations.

\acknowledgments
This research was supported by the Division of Nuclear
Physics, U.S. Department of Energy under Contract No.
DE-AC05-84OR21400  managed by
Martin Marietta Energy Systems, Inc.
 The  authors
would like to thank Dr. T. Barnes, Prof. H. Crater,  Dr. G. R. Satchler,
Prof. S. Willenbrock,
and Prof. Jian-shi Wu for helpful  discussions.
One of us (LC) would like to thank Drs. F. Plasil and M. Strayer
of Oak Ridge National Laboratory for
their kind hospitality,
and University Grants Commission of India  for partial
support.

\vfill\eject

\vfill\eject

\begin{figure}
\caption{The vertex correction factor $K^{(i)}$ for the annihilation
of a $q \bar q$ pair of various flavors.  The upper three curves are
for  color-singlet states and the lower three curves for
color-octet states.}
\end{figure}

\begin{figure}
\caption{ The vertex correction factor $K^{(f)}$ for the production of
a $q \bar q$ pair of various flavors.  The upper three curves are for
color-singlet state and the lower three curves for color-octet
states.}
\end{figure}

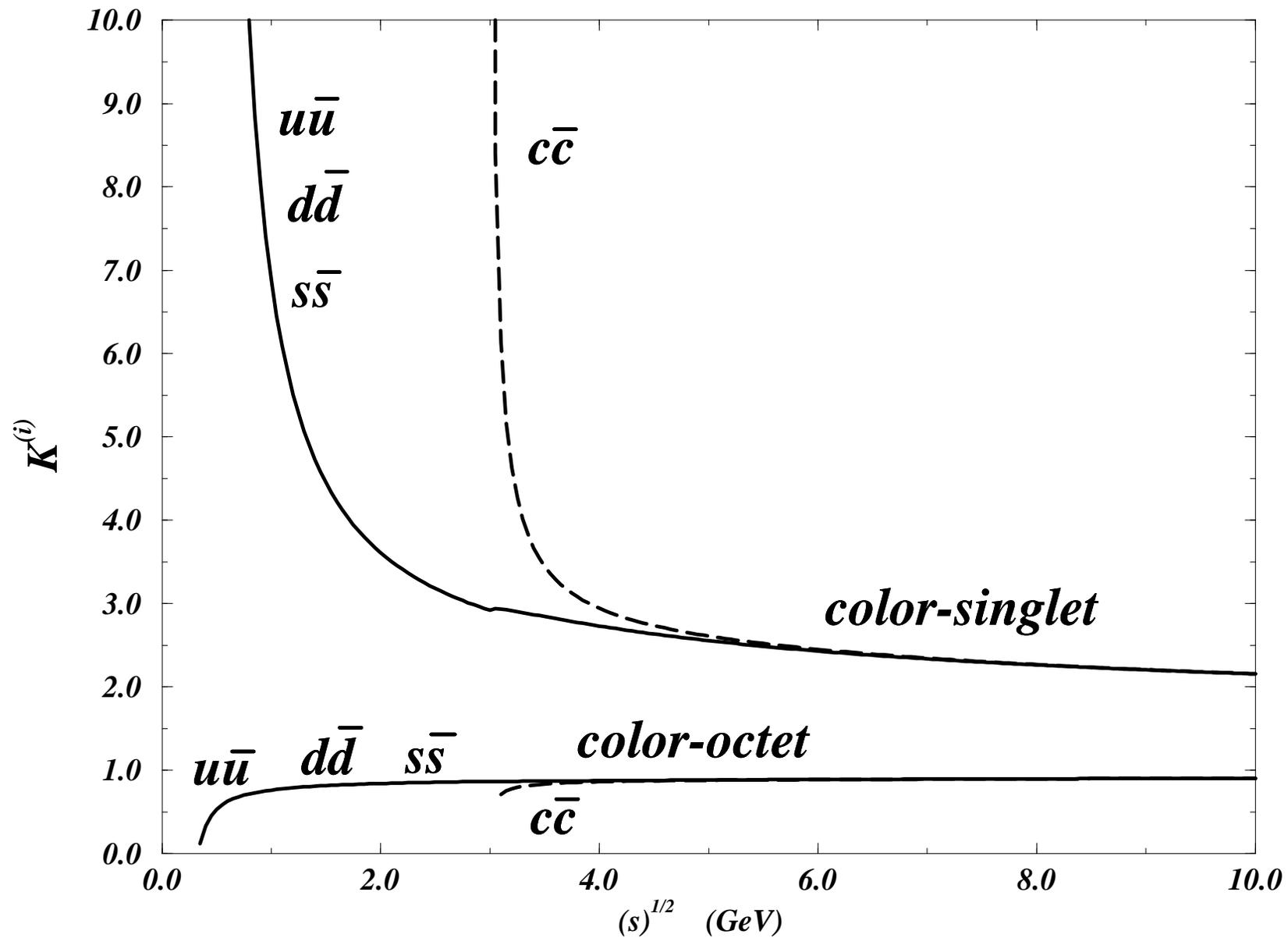
\end{document}